%% file: R1_manuscript.tex
\documentclass[twocolumn]{aastex7}

\begin{document}

%\title{Constraints on the progenitor and explosion of SN~2024ggi from hydrodynamic simulations}
\title{Constraints on the progenitor and explosion of SN~2024ggi in harmony with pre-explosion detection and hydrodynamic simulations}

\author[0000-0002-9928-0369]{Amar~Aryan}
\email[]{amararyan941@gmail.com, amar@astro.ncu.edu.tw}  
\affiliation{Graduate Institute of Astronomy, National Central University, 300 Jhongda Road, 32001 Jhongli, Taiwan}

\author[0000-0003-2284-4469]{Erin~Higgins}
\email[]{e.higgins@qub.ac.uk}  
\affiliation{Astrophysics Research Centre, School of Mathematics and Physics, Queens University Belfast, Belfast BT7 1NN, UK}

\author[0000-0002-2555-3192]{Matt~Nicholl}
\email[]{matt.nicholl@qub.ac.uk}
\affiliation{Astrophysics Research Centre, School of Mathematics and Physics, Queens University Belfast, Belfast BT7 1NN, UK}

\author[0000-0002-1066-6098]{Ting-Wan~Chen}
\email[]{twchen@astro.ncu.edu.tw}  
\affiliation{Graduate Institute of Astronomy, National Central University, 300 Jhongda Road, 32001 Jhongli, Taiwan}

\author[0009-0005-2378-2601]{Yu-Hsuan~Liu}
\email[]{m1129010@astro.ncu.edu.tw}  
\affiliation{Graduate Institute of Astronomy, National Central University, 300 Jhongda Road, 32001 Jhongli, Taiwan}

\correspondingauthor{A.~Aryan}
%\email[show]{amar@astro.ncu.edu.tw}
\email[show]{amararyan941@gmail.com, amar@astro.ncu.edu.tw}  

\correspondingauthor{T.-W.~Chen}
%\email[show]{amar@astro.ncu.edu.tw}
\email[show]{twchen@astro.ncu.edu.tw}  
%\altaffiliation{Kitt Peak National Observatory}

%\collaboration{all}{The Terra Mater collaboration}

\begin{abstract}

Supernova (SN) 2024ggi is a nearby Type II SN discovered by ATLAS, showing early flash-ionization features. The pre-explosion images reveal a red supergiant (RSG) progenitor with an initial mass of 10–17\,M$_\odot$. In the present work, we perform detailed hydrodynamic modeling to refine and put robust constraints on the progenitor and explosion parameters of SN~2024ggi. Among the progenitor models in our study, the pre-SN properties of the 11\,M$_{\odot}$ match the pre-explosion detected progenitor well. However, we find it difficult to completely rule out the 10\,M$_{\odot}$ and 12\,M$_{\odot}$ models. Thus, we provide a constraint of 11$^{+1}_{-1}$\,M$_{\odot}$ on the initial mass of the progenitor. To match the observed bolometric light curve and velocity evolution of SN~2024ggi, the favored model with an initial mass of 11\,M$_{\odot}$ has a pre-SN radius of 800\,R$_{\odot}$. This model requires an explosion energy of [0.7-0.8]$\times$10$^{51}$\,erg, nickel mass of {\bf 0.049\,M$_{\odot}$}, ejecta mass of 9.1\,M$_{\odot}$, and an amount of $\sim$\,0.5\,M$_{\odot}$ of steady-wind CSM extended up to $\sim1.2\times10^{14}$\,cm resulting from an eruptive mass-loss rate of 1.0\,M$_{\odot}$\,yr$^{-1}$. We also incorporate the accelerated-wind CSM scenario which suggests a mass-loss rate of 1.0$\times10^{-2}$\,M$_{\odot}$\,yr$^{-1}$ and a CSM mass of $\sim$\,0.7\,M$_{\odot}$ extended up to $\sim1.1\times10^{14}$\,cm. This mass-loss rate falls within the range constrained observationally.  Additionally, due to the constraint of 11$^{+1}_{-1}$\,M$_{\odot}$ on the initial mass, the range of pre-SN radius and ejecta mass would be [690--900]\,R$_{\odot}$, and [8.2--9.6]\,M$_{\odot}$, respectively.

\end{abstract}



\section{Introduction} 

Followed by the discovery of the core-collapse supernova (CCSN) 2023ixf \citep[][]{2023TNSTR1158....1I} at a distance less than 10 Mpc (DLT10), another DLT10 CCSN~2024ggi was discovered in NGC~3621 by the Asteroid Terrestrial-impact Last Alert System (ATLAS) on the 11$^{\rm th}$ of April 2024 \citep[JD=2460411.64069,][]{2024TNSTR1020....1T,2024TNSAN.100....1S,2025ApJ...983...86C} and it was classified as Type II \citep[][]{2024TNSAN.103....1H,2024TNSCR1031....1Z}. Immediate spectroscopic follow-up starting only a few hours since discovery revealed the photo-ionized, optically-thick CSM surrounding the exploding star through the presence of strong, flash-ionization features in the spectra \citep[][]{2024ApJ...972..177J,2024A&A...688L..28P,2024ApJ...972L..15S,2024ApJ...970L..18Z,2025ApJ...983...86C}.

Due to its proximity, the site of SN~2024ggi had been visited by numerous telescopes on several occasions before the explosion. Utilizing the pre-explosion images from the Hubble Space Telescope (HST) and the {\textit{Spitzer}} Space Telescope, \citet[][]{2024ApJ...969L..15X} identified a red, bright variable star with varying absolute $F814W$-band magnitudes of $-6.2$\,mag in 1995 to $-7.2$\,mag in 2003. The variation in $F814W$-band magnitudes was found consistent with a normal RSG star. For the identified progenitor, \citet[][]{2024ApJ...969L..15X} constrained the temperature and bolometric luminosity to be $T_{*}$ = $3290^{+19}_{-27}$\,K and log($L$/L$_{\odot}$) = $4.92^{+0.05}_{-0.04}$, respectively. 

In another study, \citet[][]{2024ApJ...971L...2C} examined the multi-band deep archival images in $griz$-bands from the Dark Energy Camera Legacy Survey, and provided the range of progenitor mass lying in 14--17\,M$_{\odot}$. However, their progenitor mass estimation could be influenced by the flux contamination from a nearby source, as \citet[][]{2024TNSAN.100....1S} had indicated that the red source was resolved into two sources in $F814W$-band image from HST. \citet[][]{2024TNSAN.105....1Y} also measured the brightness at the location of SN~2024ggi in the DESI Legacy Survey images and found absolute magnitudes within the range typical for an RSG. Further utilizing the HST images,  \citet[][]{2024ApJ...977L..50H} put constraints on the progenitor mass through independent measurements unaffected by the circumstellar extinction and pulsational brightness variability. They found the dying star associated with the youngest population in SN environment with an age (log[$t$/yr]) of 7.41, corresponding to an initial mass of 10.2$^{+0.06}_{-0.09}$\,M$_{\odot}$. 

Despite some constraints over the progenitor masses through the direct detection of the possible progenitor in pre-explosion images, detailed hydrodynamic simulations are {\bf scarce} in the literature to strengthen the claims on the progenitor star and explosion properties. However, \citet[][]{2025arXiv250301577E} utilized a grid of progenitor models available from \citet[][]{1988PhR...163...13N} to reproduce the observed bolometric light curve and velocity evolution of SN~2024ggi. They constrained an initial progenitor mass of 15\,M$_{\odot}$. 

In the present study, we performed detailed hydrodynamic simulations of the possible progenitor models and constrained the explosion properties of the exploding star. The rest of the paper is organized as follows. Section~\ref{sec:Sec2} details the numerical setups to generate several progenitor models with different zero-age main-sequence (ZAMS) masses and evolve them up to the pre-SN stages. Section~\ref{sec:Sec3} details the explosion setups to mimic the SN~2024ggi properties, thus putting important constraints on the explosion and circumstellar properties. Finally, in Section~\ref{sec:Sec4}, we summarize our key findings and implications of our analysis on putting the constraints over the progenitor and explosion properties of SN~2024ggi.

\section{Constraints on the possible progenitor} \label{sec:Sec2}
As seen above, initially, the direct detection of the possible progenitor in the pre-explosion images put a constraint of 13 -- 17\,M$_\odot$ \citep[][]{2024ApJ...969L..15X,2024ApJ...971L...2C} on the progenitor mass. However, an environmental analysis (free from the effect of circumstellar extinction and pulsational brightness variability) by \citet[][]{2024ApJ...977L..50H} utilizing the pre-explosion HST images of SN~2024ggi site, places the progenitor towards the slightly lower mass regime. As a result, we choose [10--17]\,M$_\odot$, ZAMS star models in our simulations to serve as the possible progenitor for SN~2024ggi.

\subsection{Details on simulation setups to create pre-SN models}
Starting from the pre-main-sequence, our models arrive on the ZAMS with specified masses and respective rotations, pass through stages of successive nuclear burning and finally reach the pre-SN stages marked by the Iron-core (Fe-core) infall. To evolve the models upto the stages of Fe-core infall, we employ the state-of-the-art simulation tool {\tt MESA}, version r24.03.1 \citep[][]{2011ApJS..192....3P, 2013ApJS..208....4P, 2015ApJS..220...15P, 2018ApJS..234...34P, 2019ApJS..243...10P, 2023ApJS..265...15J}. Our models have solar metallicity (Z = 0.0142, following \citealt[][]{2009ARA&A..47..481A}) and nominal rotation, which is only the 5\% of their critical angular rotational velocity. The solar metallicity composition for SN~2024ggi progenitor is supported by \citet[][]{2024ApJ...972..177J, 2024ApJ...969L..15X,2024ApJ...971L...2C, 2025ApJ...983...86C,2024ApJ...977L..50H}. The nominal angular rotation velocity of only 5\% of their critical angular rotational velocity is supported by the fact that the progenitors of Type IIP SNe are RSGs (confirmed through direct detection of progenitors, e.g., \citealt[][]{2009ARA&A..47...63S,VanDyk2017}) that are typically very slow or non rotating due to their extended envelope \citep[][]{1998A&A...334..210H,2001A&A...373..555M}.

\begin{figure}
    \includegraphics[width=\columnwidth,angle=0]{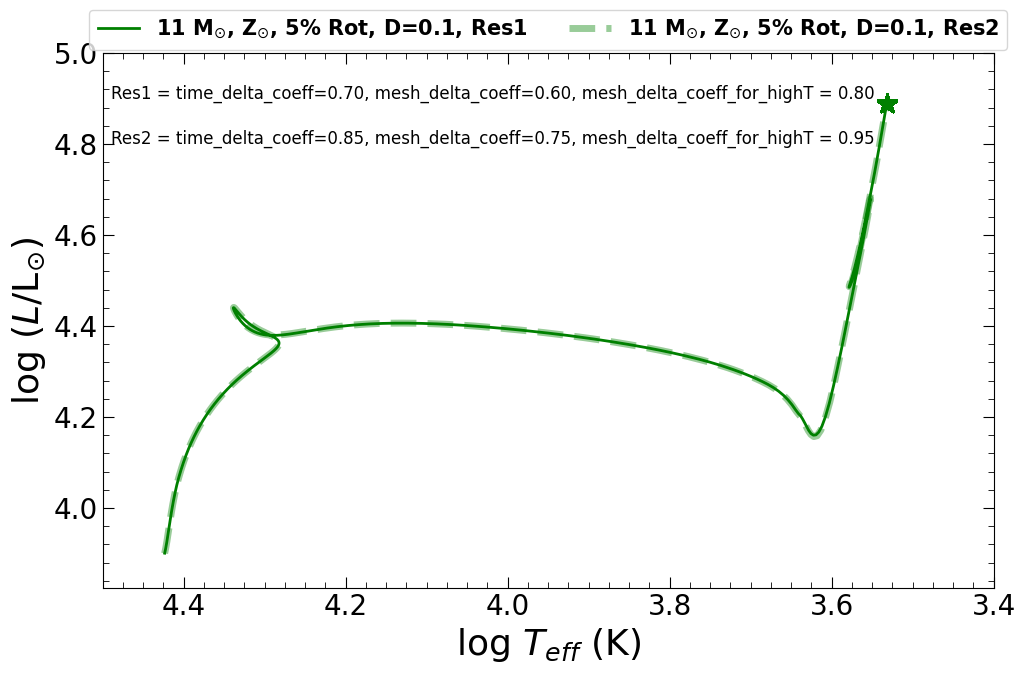}
\caption{ The stellar evolutionary tracks (from ZAMS to pre-SN stage) of 11\,M$_{\odot}$ progenitor models in our simulations, assuming two sets of temporal and spatial resolutions. The models seem to have converged for the chosen temporal and spatial resolutions.}
\label{fig:hr_resolution_test}
\end{figure}

We use the {\tt 12M\_pre\_ms\_to\_core\_collapse} test-suite directory of {\tt MESA} to create the pre-SN models. As specified in the recent versions of {\tt MESA} setups, when the hydrogen mass fraction in the core decreases to 0.68, the model is assumed to have reached the ZAMS stage (previously, the ratio of energy produced from nuclear reactions and the overall luminosity of the model being $\sim$~0.4--0.8 was used to define the arrival on ZAMS). In our models, convection is modeled following the mixing theory of \citet[][]{1965ApJ...142..841H}, adopting the Ledoux criterion. We have adopted a mixing-length-theory parameter ($\alpha_{\rm MLT}$) = 2.0 \citep[typical value being $\sim$\,1--3, as used in][]{2016ApJS..227...22F,2021MNRAS.505.2530A,2021MNRAS.500.1889O,2021MNRAS.507.1229P,2022MNRAS.517.1750A,2022JApA...43....2A,2022JApA...43...87A,2023MNRAS.521L..17A,2024ApJ...971..163R}. Following \citet[][]{1985A&A...145..179L}, semi-convection is incorporated in our models with an efficiency parameter of $\alpha_{\mathrm{sc}}$ = 0.01. To model the convective overshooting, we use the default {\tt MESA} settings for massive stars by setting overshooting parameters, $f = 0.345$ and $f_{0} = 0.01$ for the hydrogen core, assuming step overshooting. The choice of step overshooting for the hydrogen core in massive stars introduces a sharp boundary extension that ultimately helps capture the increased main-sequence lifetime and observed main-sequence width. The chosen value of $f$ and $f_{0}$ using step overshooting for hydrogen core are supported by \citet[][]{2011A&A...530A.115B} and also by asteroseismology results around this mass range \citep[][]{2020FrASS...7...70B}. The helium core would be relatively smaller, less vigorous, and the overshoot-induced mixing is expected to decay with distance. Thus, we use $f = 0.01$ and $f_{0} = 0.005$ for helium core assuming exponential overshooting utilizing the results of \citet[][]{2000A&A...360..952H}. In addition to these convective core overshooting, we also use a small amount of overshooting on the top of any other convective core to avoid spurious numerical behavior. The RSGs retain most of their outer envelope; thus, we use the `Dutch' wind scheme \citep[][]{2009A&A...497..255G} with scaling factor ($\eta$) of 0.1 only, which accounts for the small mass losses. This wind scheme incorporates the outcomes from multiple works for different situations; (a) in the regions with the effective temperature, $T_{\rm eff} > 10^4$\,K along with hydrogen mass fraction $>$\,0.4, the outcomes of \citet[][]{2001A&A...369..574V} are used; (b) when $T_{\rm eff} > 10^4$\,K and hydrogen mass fraction $<$\,0.4, the results of \citet[][]{2000A&A...360..227N} are used; and (c) the wind scheme presented in \citet[][]{1988A&AS...72..259D} is used for the regions with $T_{\rm eff} < 10^4$\,K. Throughout the simulation, we use appropriate temporal and spatial resolution. Figure~\ref{fig:hr_resolution_test} displays the effect of changing the temporal and spatial resolutions for the 11\,M$_{\odot}$, thus confirming the convergence of our model grid within the assumed resolutions in our study.

With these settings, we evolve the models up to the pre-SN stages, marked by the infall velocity of any region inside the Fe-core exceeding 1000\,km s$^{-1}$. In other words, the pre-SN stages in our simulations mark the onset of core collapse.

\begin{figure*}
\centering
    \includegraphics[width=\textwidth,angle=0]{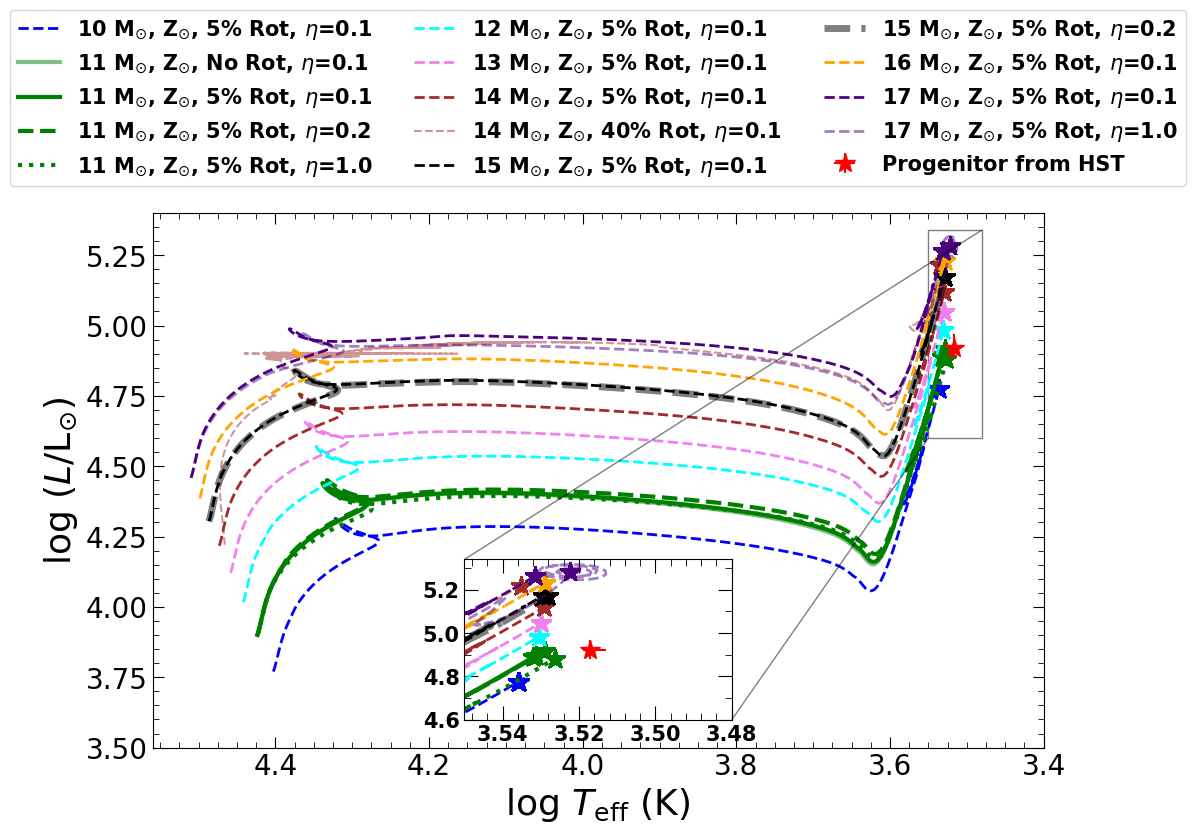}
\caption{The evolutionary tracks of 10, 11, 12, 13, 14, 15, 16, and 17 M$_{\odot}$ progenitor models in our simulations. The tracks shown here start from ZAMS and terminate at the pre-SN stages marked by $\bigstar$ for each model. The location of the HST-detected progenitor candidate is also indicated \citep[from][]{2024ApJ...969L..15X}. The inset shows our models' zoomed late stages of evolution.}
\label{fig:hr_diagram}
\end{figure*}

\subsection{Discussion on the choice of $\eta$ and rotation}
\label{choice_of_eta_n_rot}
The progenitors of Type IIP SNe are RSGs with their outer hydrogen envelope almost intact. We have assumed a small $\eta=0.1$ to account for the low mass-loss rates. There are examples \citep[e.g.,][]{2018ApJS..234...34P} where an $\eta=0.1$ has helped to produce the desired amounts of hydrogen and ejecta masses to explain the observed plateau durations of Type IIP SNe. In Figure~\ref{fig:hr_diagram}, we see that changing $\eta$ from 0.1 to 0.2 for the 11\,M$_{\odot}$ (and also 15\,M$_{\odot}$) has a negligible effect on the overall evolution of the star. Their physical properties at different stages of evolution are also almost identical, as presented in Table~\ref{tab:MESA_MODELS}. Additionally, it is argued that RSG mass-loss rates may be much lower than the `Dutch' prescription \citep[][]{2020MNRAS.492.5994B}. {\bf We also refer to \citet[][]{2025Galax..13...72V} for a review on RSG mass losses, which supports the typically modest rates of mass loss for progenitors of SNe IIP, compared to those of more massive red supergiants.} Further, while modeling the light curves of SN~2004A, SN~2004et, SN~2009ib, SN~2017eaw, and SN~2017gmr,  \citet[][]{2020ApJ...895L..45G} found several models with $\eta=0.2$ explaining the bolometric light curves and velocity evolution nicely. We also investigate the effect of changing $\eta$ to 1. One important thing to notice in Figure~\ref{fig:hr_diagram} is that the terminating positions of the models with different $\eta$ on the HR-diagram do not change considerably, as shown by the 11\,M$_{\odot}$ and 17\,M$_{\odot}$ models. However, due to an increased $\eta$, the pre-SN mass changes considerably as presented in Table~\ref{tab:MESA_MODELS}.

Rotation in stars is a natural consequence of angular momentum conservation during the gravitational collapse of a star-forming molecular cloud. Virtually all stars rotate, although the rotation rate varies widely depending on physical characteristics like mass, age, metallicity, magnetic braking, and evolutionary stage. Several studies have shown that including rotation in stellar evolution models helps to explain the observed surface chemical enrichment, extended lifetimes, position on HR-diagram, and pre-SN core masses \citep[among several others,][]{2000ARA&A..38..143M,2008A&A...479..541H,2013A&A...558A.103G}. As indicated by \citet[][]{2012A&A...537A.146E}, for the stars with initial mass range in 9--20\,M$_{\odot}$, the surface rotation velocities after the end of their respective core carbon burning stages are virtually close to zero. In the present study, we choose an initial angular rotational velocity of only 5\% of the critical angular rotational velocity. As expected, the final surface velocities in our models are exceptionally low ($\sim$\,0.02 km s$^{-1}$ for the 11\,M$_{\odot}$). We also show the non-rotating model for the 11\,M$_{\odot}$ and find that its evolution is almost identical to the slow rotating case (i.e., 5\% Rot case) within the limits of the considered temporal and spatial resolutions here. One can argue for choosing high initial angular rotational velocities, however, as shown for the 14\,M$_{\odot}$ in Figure~\ref{fig:hr_diagram}, increasing the initial angular rotational velocity from 5\% to 40\% significantly alters the overall evolution of the models. A critical consequence of increasing the initial rotation would be that the terminating stages of the models with high initial masses would move further away on the HR-diagram from the pre-explosion detected progenitor. Thus, a rapidly rotating high-mass progenitor would be less favored for SN~2024ggi.  
%($\gtrsim$13\,M$_{\odot}$)
\input{Table1}

\subsection{Constrain on the physical properties of the possible progenitor}

The evolutionary tracks of the models shown in Figure~\ref{fig:hr_diagram} start from ZAMS and terminate at their respective pre-SN stages marked by $\bigstar$ for each model. The position of the {\bf possible} progenitor detected in the pre-explosion images from HST \citep[results from][]{2024ApJ...969L..15X} is also indicated. The effective temperature $T_{\mathrm {eff}}$, and luminosity log($L$/L$_{\odot}$) of the 11\,M$_{\odot}$ model at pre-SN stage seems to align well with the progenitor candidate directly detected from HST in archival images. At the pre-SN stage, the 11\,M$_{\odot}$ model has a radius of 800\,R$_{\odot}$. Additionally, the age (log[$t$/yr]) of the 11\,M$_{\odot}$ model at the pre-SN stage is 7.38, matching closely with \citet[][]{2024ApJ...977L..50H}.

Although the pre-SN properties of 11\,M$_{\odot}$ model seems to match the physical properties of the {\bf possible} progenitor detected in pre-explosion images, it is difficult to completely rule out the 10\,M$_{\odot}$ and 12\,M$_{\odot}$ models. These models have pre-SN radii of 690\,R$_{\odot}$ and 900\,R$_{\odot}$, respectively. Thus, for the progenitor of SN~2024ggi, we provide a constrain of 11$^{+1}_{-1}$\,M$_{\odot}$ on the initial mass, and a range of pre-SN radius of 690\,R$_{\odot}$--900\,R$_{\odot}$. These constraints align nicely with the values obtained by analyzing the pre-explosion images \citep[as in][]{2024ApJ...969L..15X,2024ApJ...977L..50H}. The error bar on our initial mass constraint represents the range of masses lying in the closest vicinity of the error bar of the pre-explosion detected progenitor on the HR-diagram. The details of the physical properties of the models at different stages are provided in Table~\ref{tab:MESA_MODELS}.

\begin{figure*}
\centering
    \includegraphics[width=\columnwidth,angle=0]{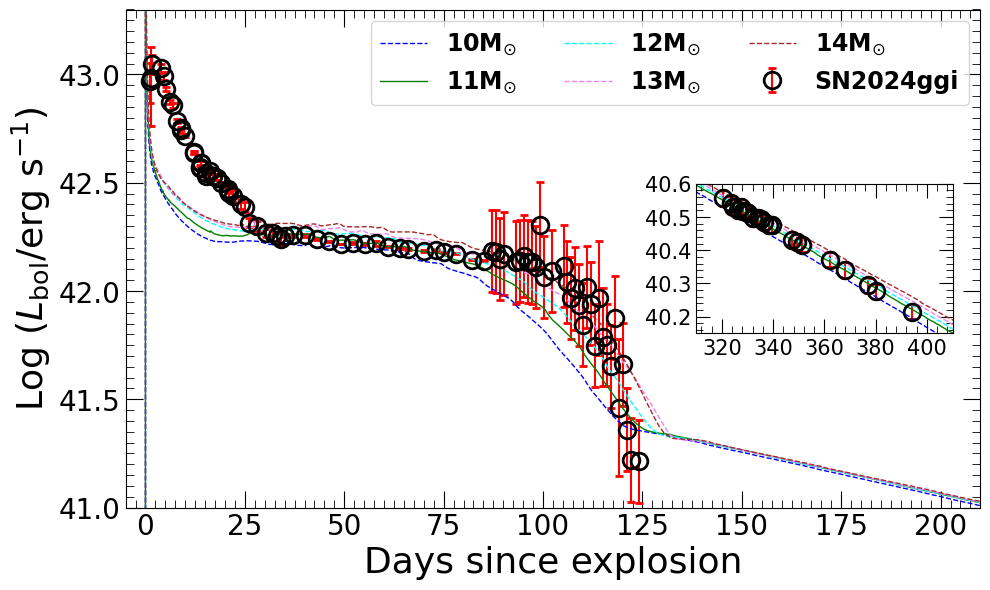}
    \includegraphics[width=\columnwidth,angle=0]{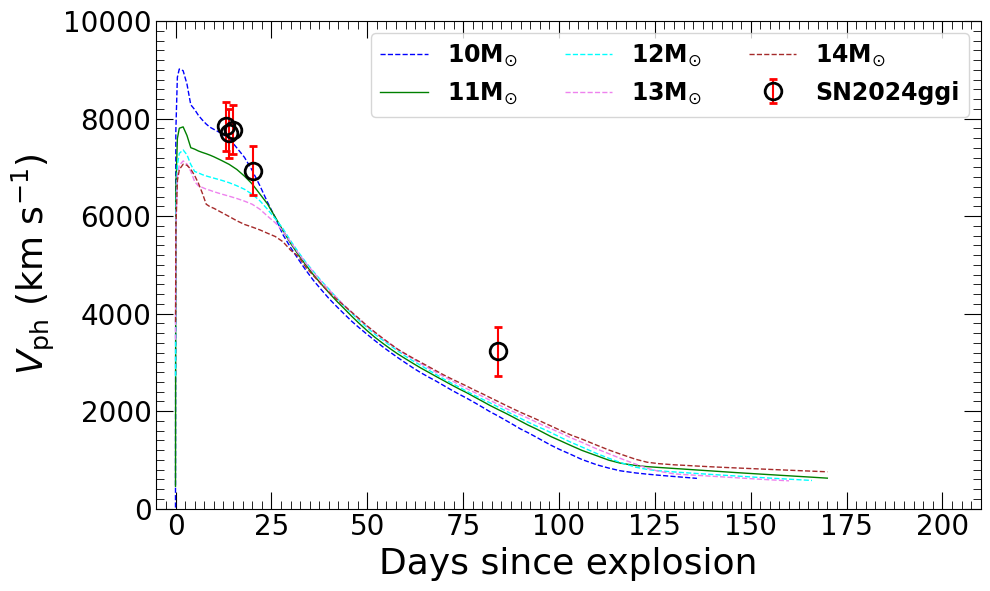}
    \includegraphics[width=\columnwidth,angle=0]{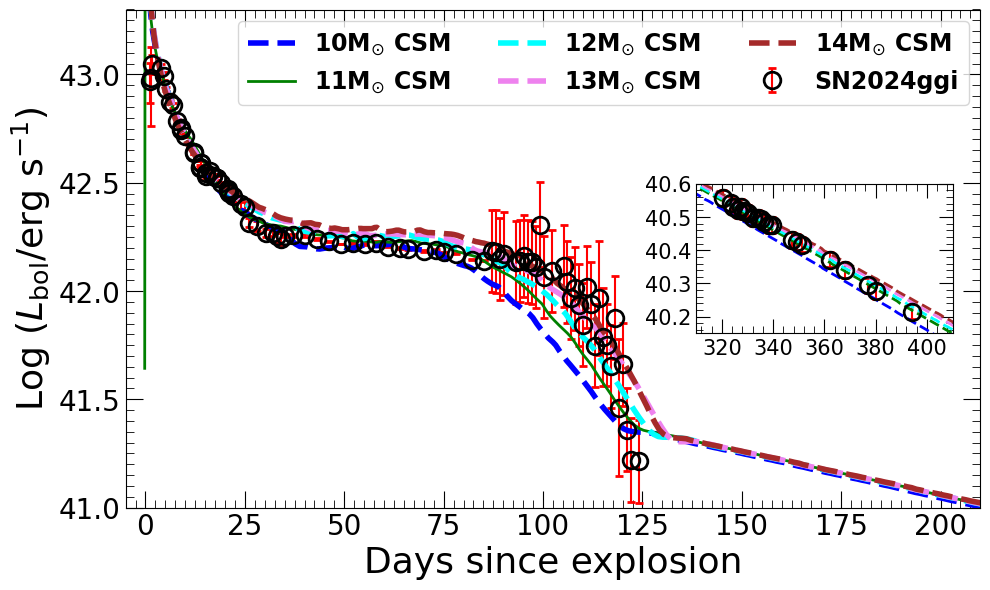}
    \includegraphics[width=\columnwidth,angle=0]{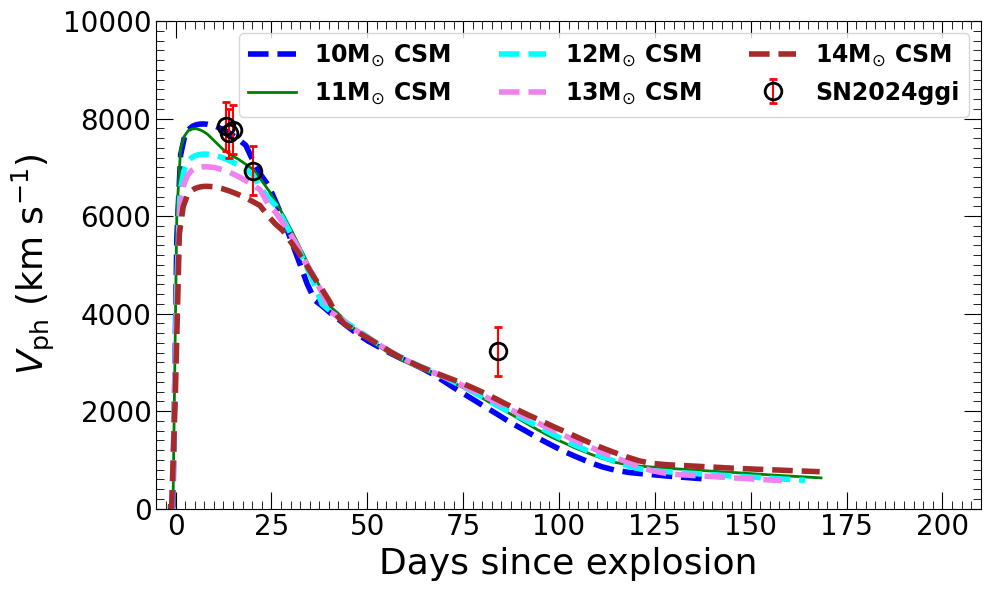}
    \includegraphics[width=\columnwidth,angle=0]{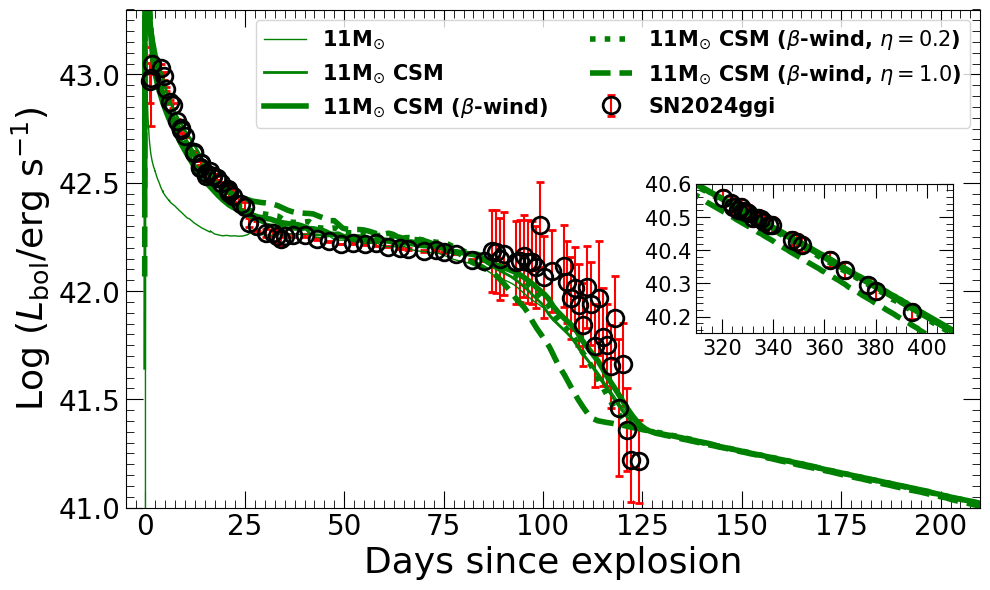}
    \includegraphics[width=\columnwidth,angle=0]{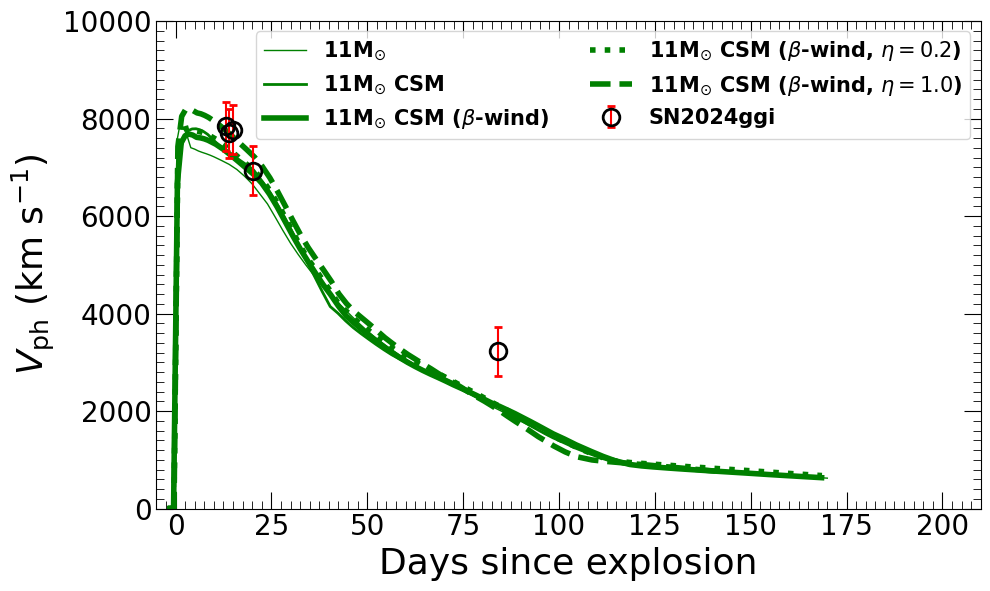}
\caption{ {\textit Top, left}: The comparison of the bolometric luminosity light curves resulting from the explosion of 10, 11, 12, 13, and 14\,M$_{\odot}$ ZAMS models with SN~2024ggi. {\textit Top, right}: Corresponding photospheric velocity evolution comparisons with the \ion{Fe}{2} line velocities of SN~2024ggi. The bolometric light curve and \ion{Fe}{2} line velocities have been imported from \citet[][]{2025arXiv250301577E}. The explosions for 14 (with 40\% rotation), 15, 16, and 17\,M$_{\odot}$ are not simulated due to their distant position on the HR-diagram from the pre-explosion detected progenitor. {\textit Middle, left}: Same as the top, left-hand panel but with a steady-state wind CSM included. {\textit Middle, right}: Corresponding photospheric velocity evolution comparisons with the \ion{Fe}{2} line velocities of SN~2024ggi. {\textit Bottom, left}: The comparison of the bolometric luminosity light curves resulting from the explosion of preferred 11\,M$_{\odot}$ model with steady state wind CSM and also with $\beta$-wind CSM. The effect of changing $\eta$ is also indicated here. {\textit Bottom, right}: Corresponding photospheric velocity evolution comparisons with the \ion{Fe}{2} line velocities of SN~2024ggi. {\bf All the models here are computed with $E_{\rm exp}=0.7\times10^{51}$ erg and $M_{\rm Ni} = 0.049$\,M$_{\odot}$ (nicely iterating the nebular phase luminosities as indicated in the insets).}}
\label{fig:explosion_no_CSM_n_with_CSM_both}
\end{figure*}

\section{Constraints on the explosion properties} \label{sec:Sec3}
In the previous section, we explored the physical properties of the models at their pre-SN stages marked by the onset of Fe-core infall. Next, the models are allowed to continue evolving until they reach near-shock breakout (SBO) stages utilizing the {\tt ccsn\_IIp} test-suite directory of {\tt MESA}. The pre-SN model from the {\tt 12M\_pre\_ms\_to\_core\_collapse} directory is incorporated as the starting point in the {\tt ccsn\_IIp} test-suite directory. In this test-suite directory, the central region of the model where the entropy per baryon reaches 4\,$k_{\rm B}$, is first removed. This excision represents the part of the model that would have collapsed to form a proto-neutron star. The model is further allowed to continue infalling until its inner boundary (IB) reaches close to the location of the stalled shock. After the first few seconds, we account for further fallback by removing negative velocity material at the IB. The explosion is simulated by injecting energy in a thin layer of approximately 0.2\,M$_{\odot}$ at the IB for 0.2\,seconds, at a rate sufficient enough to raise the total energy of the model to our specified value of $E_{\rm exp}$. We also set the amount of nickel ($M_{\rm Ni}$) synthesized in the SN. The mass of the SN ejecta ($M_{\rm ej}$) is determined by subtracting the final core mass after fallback ($M_{\rm cf}$) from the mass of the progenitor model at pre-SN stage ($M_{\rm pre-SN}$). Thus, $M_{\rm ej}$ = $M_{\rm pre-SN}$ - $M_{\rm cf}$. The detailed explanation of how the models are simulated from Fe-core infall till near SBO stages is documented in Section 6.1 of \citet[][]{2018ApJS..234...34P}.

\subsection{Constraints on $E_{\rm exp}$, $M_{\rm Ni}$, and $M_{\rm ej}$}
Once the models have reached the near SBO stages with desired $E_{\rm exp}$ and $M_{\rm Ni}$, we utilize the radiation hydrodynamics code, {\tt STELLA} \citep[][]{1998ApJ...496..454B,2000ApJ...532.1132B,2004Ap&SS.290...13B,2006A&A...453..229B} to compute the bolometric light curve and photospheric velocity evolution. {\tt STELLA} numerically solves the radiative transfer equations by employing the intensity moment approximation in each frequency bin. In typical {\tt MESA} applications, it operates with 40 frequency groups, sufficient to generate the spectral energy distributions.

The left-hand side of the top panel of Figure~\ref{fig:explosion_no_CSM_n_with_CSM_both} shows the comparison of the {\tt STELLA}-generated bolometric luminosity light curves from the synthetic explosions of our models with the observed bolometric luminosity light curve, while the top right-hand panel shows the comparison of the {\tt STELLA}-generated photospheric velocity ($V_{\rm ph}$) evolution with the observed \ion{Fe}{2} (5169 \AA)-line velocities of SN~2024ggi. The $V_{\rm ph}$ estimated by {\tt STELLA} at a particular instant is the velocity at the location of the photosphere where the Rosseland optical depth ($\tau_{\rm Ros}$) is 2/3 \citep[][]{2018ApJS..234...34P}. The observed bolometric light curve and \ion{Fe}{2}-line velocities are adopted from \citet[][]{2025arXiv250301577E}. The most plausible progenitor having an initial mass of 11\,M$_{\odot}$ requires $E_{\rm exp}$ = 0.7$\times$10$^{51}$\,erg and {\bf {\bf $M_{\rm Ni}$ = 0.049\,M$_{\odot}$} to iterate the observed bolometric light curve (barring the initial $\sim$\,30 days, details about it in the next section) and \ion{Fe}{2} velocities. We provide a discussion on the choice of this nickel mass and the consequence of choosing a lower or higher nickel mass in the next section.} A recent study by \citet[][]{2025arXiv250705803D} indicates a slightly larger $M_{\rm Ni}$ of 0.06\,M$_{\odot}$. However, it is close to the amount (0.05--0.06\,M$_{\odot}$) obtained by \citet[][]{2025arXiv250722794F}. The $E_{\rm exp}$ in our modeling seems to be smaller than \citet[][]{2024ApJ...972..177J,2025arXiv250301577E}; however, it is in line with the delayed neutrino explosion mechanism of CCSNe as investigated in \citet[][]{2024OJAp....7E..69S}.

In these two subplots, we also see the corresponding results for the 10, 12, 13, and 14\,M$_{\odot}$ models with similar $E_{\rm exp}$ and $M_{\rm Ni}$. Although all the models seem to explain the observed plateau and the falling tail of SN~2024ggi, only the 10 and 12\,M$_{\odot}$ models could attain the initial observed \ion{Fe}{2} velocities. Thus, our constraint of 11$^{+1}_{-1}$\,M$_{\odot}$ on initial progenitor mass further solidifies. The corresponding values of $M_{\rm ej}$ from the 10, 11, and 12\,M$_{\odot}$ are 8.2\,M$_{\odot}$, 9.1\,M$_{\odot}$, and 9.6\,M$_{\odot}$, respectively. Thus, we constrain the range of $M_{\rm ej}$ for SN~2024ggi to be [8.2 -- 9.6]\,M$_{\odot}$. 

Now, more than a year since the discovery of SN~2024ggi, a few nebular phase spectroscopic observations are also there in the literature. Comparing the molecule-free radiative-transfer model-generated spectra with nebular phase spectroscopy, \citet[][]{2025arXiv250705803D} constrain an explosion energy and nickel mass of $\sim$\,10$^{51}$\,erg and 0.06\,M$_{\odot}$, respectively. Further, while studying the late-time bolometric light curve, \citet[][]{2025arXiv250722794F} propose a nickel mass of 0.05 -- 0.06\,M$_{\odot}$ for SN~2024ggi. The comparisons of the [\ion{O}{1}]/[\ion{Ca}{2}] flux ratio in the nebular phase spectra with models from \citet[][]{2014MNRAS.439.3694J} and \citet[][]{2021A&A...652A..64D} reveal a progenitor having initial mass of around 10--12\,M$_{\odot}$ \citep[][]{2025arXiv250722794F}. These observational constraints on nickel mass and initial mass are well in line with our results derived from simulations.

\begin{figure*}
\centering
    \includegraphics[width=\columnwidth,angle=0]{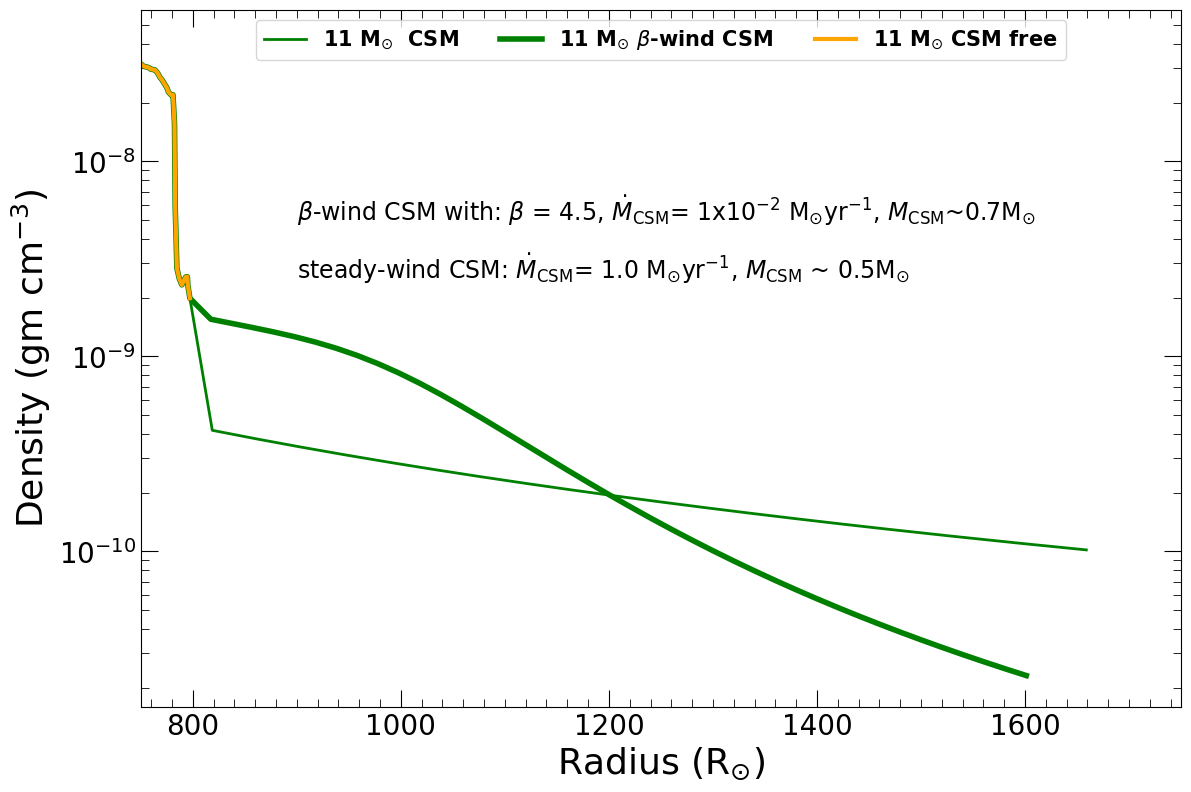}
    \includegraphics[width=\columnwidth,angle=0]{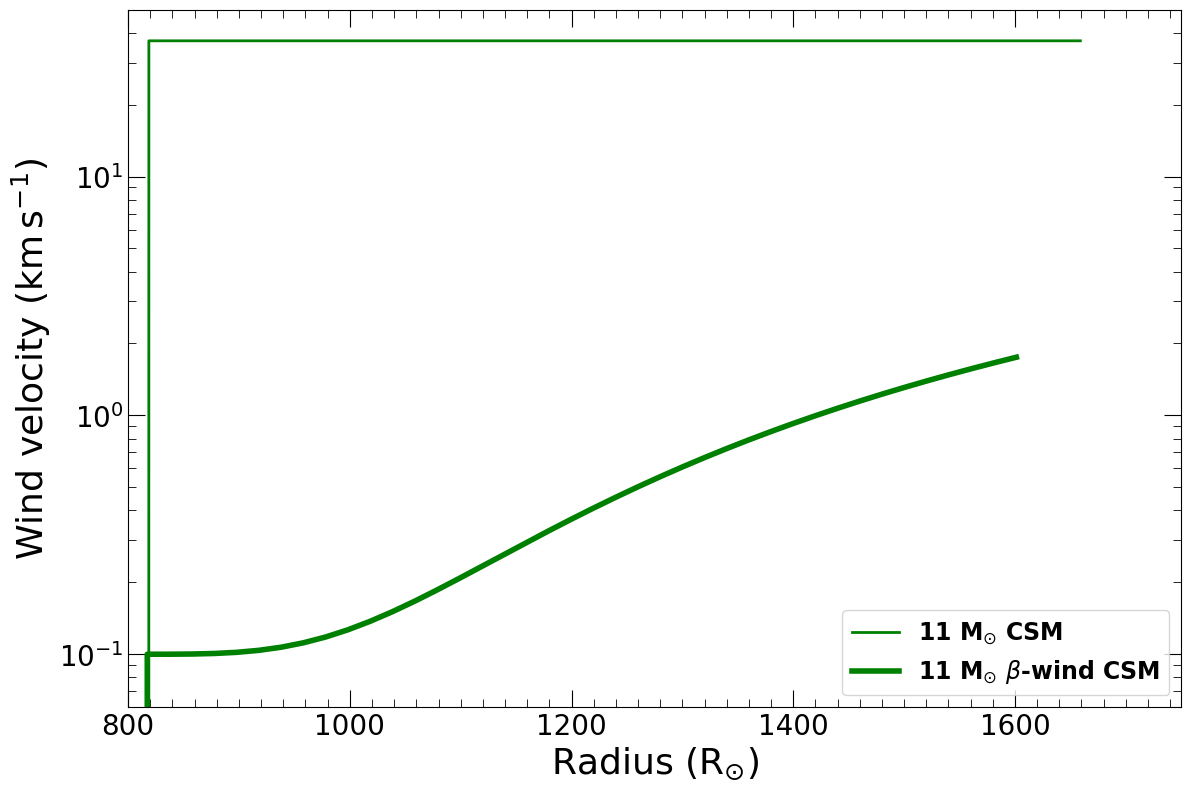}
\caption{{\textit Left}: The density distribution as a function of radius for the two different CSM models assumed in the study. {\textit Right}: Corresponding wind velocity distribution.}
\label{fig:csm_structure}
\end{figure*}

\begin{figure*}
\centering
    \includegraphics[width=0.8\textwidth,angle=0]{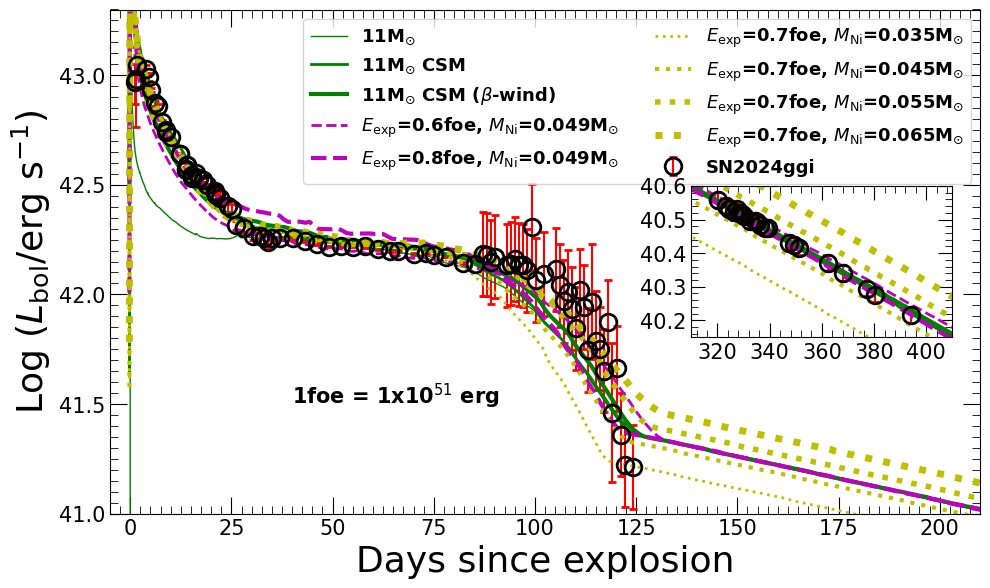}
\caption{ Effect of varying explosion energy and nickel mass on the 11\,M$_{\odot}$ model with accelerated-wind  CSM (for comparison, the steady-wind CSM model is also shown). The variations help us to constrain the range of plausible explosion energy and the amount of nickel mass.}% {\bf {\textit Right}: Results of minor variations in nickel mass to match the nebular phase luminosity. SN~2024ggi seems to have synthesized a nickel mass of 0.049\,M$_{\odot}$.}}
\label{fig:explosion}
\end{figure*}

\subsection{Constraints on the CSM properties {\bf and the effect of variation of $E_{\rm exp}$ and $M_{\rm Ni}$}}
We see in the previous section that the bolometric luminosities of the initial few days ($\sim$\,30 days) can not be reproduced by any of the models. An ad-hoc mechanism of CSM interaction has been proposed to account for the additional luminosities during the initial days of such Type II SNe. The presence of CSM around the SN~2024ggi progenitor has been confirmed by several observations \citep[][]{2024ApJ...972..177J,2024A&A...688L..28P,2024ApJ...972L..15S,2024ApJ...970L..18Z,2025ApJ...983...86C}. 

To model the CSM around the possible progenitor of SN~2024ggi, we assume steady state wind with velocity, $v_{\rm CSM}$ = 37\,km\,s$^{-1}$. The choice of this velocity is supported by the value of 37$\pm$4\,km\,s$^{-1}$ obtained by \citet[][]{2024ApJ...972L..15S} using high-resolution spectroscopic observations during the early phases of SN~2024ggi. For all the models with CSM, we incorporate the wind duration, $t_{\rm CSM}$ = 0.5\,yr and mass-loss rate, $\dot{M}_{\rm CSM}$ = 1.0\,M$_{\odot}$\,yr$^{-1}$. With the assumed mass-loss rate, the amount of CSM mass ($M_{\rm CSM}$) around our progenitor models is $\sim$\,0.5\,M$_{\odot}$, which is slightly higher than \citet[][]{2025ApJ...983...86C} but similar to \citet[][]{2025arXiv250301577E}. For our preferred model having initial mass of 11\,M$_{\odot}$, the extent of CSM reaches up to 1.2$\times$10$^{14}$\,cm ($\sim$\,1660\,R$_{\odot}$). The inferred mass-loss indicates that the progenitor suffered an enhanced mass-loss during the last $\sim$\,180 days before the explosion. We find that the $\dot{M}_{\rm CSM}$ in our modeling to produce the right amount of CSM responsible for ejecta-CSM interaction is smaller than $\dot{M}_{\rm CSM}$ = 3.6\,M$_{\odot}$\,yr$^{-1}$ obtained by \citet[][]{2025arXiv250301577E} in the steady wind case. The explosion results with steady-wind CSM are shown in the middle panels of Figure~\ref{fig:explosion_no_CSM_n_with_CSM_both}. It may not be very likely for any RSG star to have such a high mass-loss rate. As a consequence, we also incorporate the accelerated-wind (also known as the $\beta$-wind) scenario as prescribed by \citet[][]{2018MNRAS.476.2840M}.

For the accelerated-wind case, we adopt the $\beta$ velocity law for the wind as mentioned in \citet[][]{2018MNRAS.476.2840M},
$v_{\rm CSM} = v_{0} + (v_{\infty} - v_{0})(1-R_{0}/r)^{\beta}$. Here, $v_{0}$ is the initial wind velocity, and it is so chosen that the CSM density is smoothly connected from the surface of the progenitor. $R_{0}$ is the radial coordinate where the CSM is attached to the progenitor model. $v_{\infty}$ is the terminal wind velocity which is 37\,km\,s$^{-1}$ from \citet[][]{2024ApJ...972L..15S}. We also use $v_{0}$ = 0.1\,km\,s$^{-1}$ (from \citealt[][]{2025arXiv250301577E}) and ${\beta}$ = 4.5 (accepted values being 1--5, as presented in \citealt[][]{2018MNRAS.476.2840M}) while modeling the initial stages of SN~2024ggi. We tried different combination of CSM properties, however, the preferred model having initial mass of 11\,M$_{\odot}$ with accelerated-wind scenario required a mass-loss rate $\dot{M}_{\rm CSM}$ = 1$\times$10$^{-2}$\,M$_{\odot}$\,yr$^{-1}$ with the CSM extended upto 1.1$\times$10$^{14}$\,cm ($\sim$\,1600\,R$_{\odot}$). The bottom panels of Figure~\ref{fig:explosion_no_CSM_n_with_CSM_both} show the corresponding comparisons of the bolometric light curve (left) and photospheric velocity evolution (right) for the $\beta$-wind CSM case. The accelerated-wind mass-loss rate in our case is in line with \citet[][]{2024ApJ...972..177J, 2024ApJ...972L..15S, 2024ApJ...970L..18Z}. An amount of $\sim$\,0.7\,M$_{\odot}$ is required to explain the initial enhancement in the luminosity. Thus, the accelerated-wind CSM structure requires a slightly higher amount and nearly similar extent of CSM to explain the early enhancement in the luminosity. From the velocity evolution panels in the second column of the Figure~\ref{fig:explosion_no_CSM_n_with_CSM_both}, we notice that at the early times, the favored models (i.e., the models with 11\,M$_{\odot}$) show little/no difference between the photospheric velocity and the \ion{Fe}{2} line velocities. However, as the photosphere moves deeper into the ejecta, we see that the two velocities start to diverge substantially and the actual measured \ion{Fe}{2} velocity point lies slightly higher than the $V_{\rm ph}$ curve \citep[a well-known behavior mentioned in][]{2018ApJS..234...34P}.

The effect of changing $\eta$ on the light curve is also shown in the bottom panels of Figure~\ref{fig:explosion_no_CSM_n_with_CSM_both}. A small increment in $\eta$ (0.1 to 0.2) has a negligible effect on the light curve shape; however, a large change (0.1 to 1.0) significantly reduces the plateau duration. Due to an increased mass-loss rate, this reduction in plateau duration is attributed to a smaller outer hydrogen envelope, which controls plateau duration. This analysis also rules out the possibility of a much lower (say, 9\,M$_{\odot}$) initial mass progenitor for SN~2024ggi, as it might not have massive enough hydrogen envelope to explain the observed plateau duration.

To provide a detailed view of the CSM assumed in the present analysis, the density structures of the steady-wind and accelerated-wind scenarios are shown in the left-hand panel of Figure~\ref{fig:csm_structure}.  The right-hand panel of Figure~\ref{fig:csm_structure} displays the wind velocity distribution for the two CSM structures. In case of $\beta$-wind CSM, the wind velocity near the progenitor star is very small, which smoothly increases as the radius increases; however, in the steady state CSM case, the wind velocity remains fixed to 37\,km\,s$^{-1}$ throughout. {\bf A recent study by \citet[][]{2025Galax..13...72V} pointed out that RSGs have extended, dense atmospheres from which only a fraction is lost through a stellar wind. The high inferred mass-loss rates in the years preceding an SN explosion likely reflect the dense atmosphere, not a wind, and therefore do not represent mass loss and should instead by quantified in terms of density only.}

From the preferred 11\,M$_{\odot}$ model with accelerated-wind CSM, we try to constrain the $E_{\rm exp}$ and $M_{\rm Ni}$ by varying one of these two parameters while keeping the other fixed. As shown in Figure~\ref{fig:explosion}, a reduced $E_{\rm exp}$ of 0.6$\times$10$^{51}$\,erg seems to slightly over produce the plateau duration in the observed bolometric luminosity light curve. If we increase the $E_{\rm exp}$ to 0.8$\times$10$^{51}$\,erg, the tail phase improves, although one can notice that the plateau luminosity is increasing. Thus, we can not vary the explosion energy much and provide a range of [0.7--0.8]$\times$10$^{51}$\,erg.  %Thus, due to the absence of a robust tail-phase follow-up observations, an upper limit of  $\lesssim$0.055\,M$_{\odot}$ is provided for $M_{\rm Ni}$. This limit on $M_{\rm Ni}$ aligns perfectly with the values quoted with recent nebular phase spectroscopy and tail phase analysis of SN~2024ggi \citep[][]{2025arXiv250722794F,2025arXiv250705803D}. 
{\bf As shown in Fig.~\ref{fig:explosion}, the model with $M_{\rm Ni}$ of 0.049\,M$_{\odot}$ nicely iterates the observed luminosities. The amount of the nickel mass absolutely matches to the value estimated by the analytical model fit represented by the equation 1 in \citet[][]{2025arXiv250722794F}. Further, we see that the models with  $M_{\rm Ni}$ = 0.055\,M$_{\odot}$ and 0.065\,M$_{\odot}$ slightly over-estimate the nebular phase luminosity. Moreover, a lesser amount (0.035\,M$_{\odot}$ and 0.045\,M$_{\odot}$) of $M_{\rm Ni}$ underestimates the nebular phase luminosity. Thus, we robustly constrain an amount of 0.049\,M$_{\odot}$ of nickel synthesized in SN~2024ggi.}

\section{Discussion and Conclusion} \label{sec:Sec4}
In the current work, we performed detailed hydrodynamic simulations for SN~2024ggi. We started by choosing a range of progenitors having initial masses of [10--17]\,M$_{\odot}$, evolved them up to their pre-SN stages, and finally simulated their explosions to mimic the observed bolometric light curve and velocity evolution of SN~2024ggi. We utilized one of the most advanced modules, {\tt MESA}, in conjunction with {\tt STELLA} to perform the entire simulations. 

\begin{itemize}

    \item Among all the models in our study, the location of 11\,M$_{\odot}$ at its pre-SN stage on the HR-diagram matched well with the directly detected progenitor {\bf candidate} in the pre-explosion images. However, we found it difficult to completely rule out the 10\,M$_{\odot}$ and 12\,M$_{\odot}$ models due to their proximity to the directly detected progenitor candidate on the HR-diagram. Thus, although the 11\,M$_{\odot}$ progenitor appeared more plausible, we provided a constrain of 11$^{+1}_{-1}$\,M$_{\odot}$ on the progenitor mass. The pre-SN radius of the preferred 11\,M$_{\odot}$ model was 800\,R$_{\odot}$, however, due to the mass constraint obtained, the range of pre-SN radius would be [690 -- 900]\,R$_{\odot}$. These constraints on mass and radius were in line with the observationally estimated values reported in \citet[][]{2024ApJ...969L..15X,2024ApJ...977L..50H}.

    \item To explain the overall observed bolometric light curve and velocity evolution, the plausible model with an initial mass of 11\,M$_{\odot}$ required an $E_{\rm exp}$ of [0.7--0.8]$\times$10$^{51}$ erg, $M_{\rm Ni}$ of {\bf 0.049\,M$_{\odot}$}, $M_{\rm ej}$ of 9.1\,M$_{\odot}$, and an amount of $\sim$\,0.7\,M$_{\odot}$ of CSM extended up to $\sim$\,1600\,R$_{\odot}$ (assuming accelerated-wind CSM). However, due to the constraint of 11$^{+1}_{-1}$\,M$_{\odot}$ on the initial mass, the range of pre-SN radius and ejecta mass would be [690--900]\,R$_{\odot}$, and [8.2--9.6]\,M$_{\odot}$.  The overall pre-SN properties and explosion results for 13 and 14\,M$_{\odot}$ models in our simulations disfavored a high mass progenitor for SN~2024ggi. %{\bf Later on, with access to nebular phase bolometric luminosities, our models estimate that SN~2024ggi synthesized 0.049\,M$_{\odot}$ of nickel mass.} 
    
    \item Comment on the initial mass degeneracy: While performing hydrodynamic modeling of the Type IIP SNe, \citet[][]{2020ApJ...895L..45G} indicated initial mass degeneracies. Different initial mass progenitors could produce similar bolometric light curves and velocity evolution from different initial physical conditions (e.g., rotation, mass-loss, etc.). However, we saw how changing the initial parameters (rotation and eta) in our models impacted their overall evolution and terminating positions on the HR-diagram. In the present study, we had additional informations on the progenitor candidate's luminosity, effective temperature, and mass through pre-explosion {\bf progenitor candidate} detection. Thus, the preferred initial mass would be the one that lied in the proximity of the position of the pre-explosion detected progenitor candidate on the HR diagram. Ultimately, our constrained progenitor with initial mass of 11$^{+1}_{-1}$\,M$_{\odot}$ not only matched the position of the pre-explosion detected progenitor {\bf candidate} on the HR-diagram, but also explained the observed bolometric light curve and velocity evolution satisfactorily.
\end{itemize}

\begin{acknowledgments}
{\bf We thank the anonymous referee for providing constructive comments helpful in improving the manuscript further.} AA thanks Prof. Takashi J. Moriya for useful communications. We are thankful to K. Ertini for kindly sharing the bolometric light curve and Fe II line velocity evolution of SN~2024ggi. {\bf We are thankful to L. Ferrari for kindly sharing the bolometric data points of the nebular stage.} AA and T.-W.C.  acknowledge the Yushan Young Fellow Program by the Ministry of Education, Taiwan for the financial support (MOE-111-YSFMS-0008-001-P1). 
T.-W.C. acknowledges funding from the National Science and Technology Council, Taiwan (NSTC grant 114-2112-M-008-021-MY3) to support GREAT Lab. EH acknowledges the Leverhulme Trust for funding (grant ECF-2024-262). MN is supported by the European Research Council (ERC) under the European Union's Horizon 2020 research and innovation programme (grant agreement No.~948381).
\end{acknowledgments}

\bibliography{manuscript}{}
\bibliographystyle{aasjournalv7}



\end{document}

%% file: Table1.tex
\begin{table*}
\caption{The ZAMS and pre-SN properties of progenitor models using {\tt MESA} along with the {\tt STELLA} explosion parameters.}
\label{tab:MESA_MODELS}
\begin{center}
{\scriptsize
\begin{tabular}{ccccccccccccc}
\hline \hline

  &  ZAMS & & \hspace{1.3cm}\vline&  &  &  Pre-SN &\hspace{1.3cm}\vline & &  Explosion\\
\hline
 $M_{\rm ZAMS}^{a}$	& $T_{\mathrm{eff}}$  & $R_{\mathrm{ZAMS}}^{b}$ & log$L_{\rm ZAMS}^{c}/{\rm L_{\odot}}$  &	$M_{\rm pre-SN}^{d}$	& $T_{\mathrm{eff}}$  & $R_{\mathrm{pre-SN}}^{e}$ & log$L_{\rm pre-SN}^{f}/{\rm L_{\odot}}$ 	&	$M_{\mathrm{cf}}^{g}$ & $M_{\mathrm{ej}}^{h}$ & $M_{\mathrm{Ni}}^{i}$ &	$E_{\mathrm{exp}}^{j}$ 	\\
	(M$_{\odot}$) & K	&	(R$_{\odot}$)	 &   &      (M$_{\odot}$) & K	&	(R$_{\odot}$)	 &   & (M$_{\odot}$) & (M$_{\odot}$) & (M$_{\odot}$) &	($10^{51}$\,erg)\\ 	
\hline
\hline

	10.0  	&	25300  &  4.0   & 3.77 & 9.9  &    3450  & 690 & 4.78  &  1.68 & 8.2 &  {\bf 0.049} & 0.7	\\

	11.0$^{\dagger}$  	&	26500  &  4.2   & 3.90  & 10.8  &    3400  & 800 & 4.89 & -- & -- &  -- & --		\\
    
	11.0  	&	26500  &  4.2   & 3.90  & 10.8  &    3400  & 800 & 4.89 & 1.71 & 9.1 &  {\bf 0.049} & 0.7		\\

	11.0$^{*}$  	&	26500  &  4.2   & 3.90  & 10.6  &    3400  & 840 & 4.92 & 1.71 & 8.9 &  {\bf 0.049} & 0.7		\\

	11.0$^{**}$  	&	26500  &  4.2   & 3.90  & 9.5  &    3400  & 820 & 4.88 & 1.68 & 7.8 &  {\bf 0.049} & 0.7		\\
    
	12.0  	&	27700  &  4.4   & 4.02 & 11.8  &    3390  & 900 & 4.98 & 2.17 & 9.6 &  {\bf 0.049}	& 0.7	\\

	13.0  	&	28700  &  4.6    & 4.12 & 12.7  &    3390  & 970 &  5.05 & 2.47 & 10.2 &  {\bf 0.049} & 0.7	\\

	14.0  	&	29700  &  4.8    & 4.22 &  13.6 &    3380  & 1060 & 5.12 & 1.95 & 11.7 &  {\bf 0.049}	& 0.7	\\

	14.0$^{\ddagger}$  	&	29300  &  5.0    & 4.22 &  13.4 &    3400  & 1150 & 5.22 & -- & -- &  --	& --	\\

	15.0  	&	30700  &  5.0    & 4.30 &  14.5 &    3380  & 1120 & 5.17 & -- & -- &  --	& --	\\

	15.0$^{*}$  	&	30700  &  5.0    & 4.30 &  14.1 &    3370  & 1130 & 5.17 & -- & -- &  --	& --	\\

	16.0  	&	31500  &  5.2    & 4.38 &  15.4 &    3380  & 1200 & 5.23 & -- & -- &  --	& --	\\

	17.0 	&	32400  &  5.4    & 4.46 &  16.5 &    3400  & 1240 & 5.27 & -- & -- &  --	& --	\\

    17.0$^{**}$ 	&	 32400  &  5.4    & 4.46 &  10.1 &    3330  & 1310 & 5.28 & -- & -- &  --	& --	\\\hline
\multicolumn{0}{>{\centering\arraybackslash}p{0.1\textwidth}}{
    \parbox{0.6\textwidth}{\centering
    Additional models: 11 M$_{\odot}$ with fixed $M_{\rm Ni}$=0.049\,M$_{\odot}$ and different $E_{\rm exp}$}
} \\
%\multicolumn{2}{|c|}{\parbox{5cm}{11 M$_{\odot}$ with fixed $M_{Ni}$=0.054\,M$_{\odot}$ erg and different $E_{exp}$}}\\
\hline
\hline
	  	&	  &     &   &   &      &  &  & 1.73 & 9.1 &  {\bf 0.049} & 0.6		\\

%     &	  	&	  &     &   &   &      &  &  & 2.00 & 22.99 &  0.056 & 0.7		\\

	  	&	  &     &   &   &     &  &  & 1.69 & 9.1 &  {\bf 0.049} & 0.8		\\
\hline

\multicolumn{0}{>{\centering\arraybackslash}p{0.1\textwidth}}{
    \parbox{0.6\textwidth}{\centering
    Additional models: 11 M$_{\odot}$ with fixed $E_{exp}$=0.7$\times$10$^{51}$ erg and different $M_{Ni}$}
} \\
%\multicolumn{2}{|c|}{\parbox{5cm}{11 M$_{\odot}$ with fixed $E_{exp}$=0.7*10$^{51}$ erg and different $M_{Ni}$}}\\
\hline
\hline
	  	&	  &     &   &   &      &  &  & 1.71 & 9.1 &  0.035 & 0.7		\\
        
	  	&	  &     &   &   &      &  &  & 1.71 & 9.1 &  0.045 & 0.7		\\

	  	&	  &     &   &   &      &  &  & {\bf 1.71} & {\bf 9.1} &  {\bf 0.049} & {\bf 0.7}		\\

        %&	  &     &   &   &      &  &  & {\bf 1.71} & {\bf 9.1} &  {\bf 0.053$^{***}$} & {\bf 0.7}		\\
                
	  	&	  &     &   &   &      &  &  & 1.71 & 9.1 &   0.055 & 0.7	\\
        
	  	&	  &     &   &   &      &  &  & 1.71 & 9.1 &  0.065 & 0.7		\\
\hline
		
\end{tabular}}
\end{center}
%\par
{$^a$ Mass at ZAMS.
$^b$Progenitor radius at ZAMS,
$^c$Luminosity at ZAMS,
$^d$Final mass at the pre-SN stage,
$^e$Pre-SN phase radius,
$^f$Pre-SN phase luminosity,
$^g$Final core mass after fallback,
$^h$Ejecta mass,
$^i$Amount of synthesized Nickel,
$^j$total energy after explosion energy.
$^\dagger$ The non-rotating case. The properties of the non-rotating 11\,M$_{\odot}$ model are indistinguishable with the slowly rotating model within the limits of resolutions considered here. It's explosion results are expected to be almost similar as the slow rotating case (i.e., similar to Rot 5\% case).
$^*$ With $\eta$ = 0.2.
$^{**}$ With $\eta$ = 1. 
$^\ddagger$ With initial angular rotation velocity of 40$\%$ of critical angular rotation velocity. 
Unless mentioned, $\eta$ = 0.1 is used. {\bf All the {\tt MESA}-inlist-files, pre-SN models, {\tt STELLA} input files to generate luminosity, velocity evolution, and also the final {\tt STELLA} result files, are available on \dataset[Zenodo, MESA community]{https://zenodo.org/records/17396113} and also on \dataset[Zenodo, AAS Journals community]{https://zenodo.org/records/17395362}.}}

\end{table*}